\documentclass[prb,draft,amsfonts,amssymb,amsmath,showpacs]{revtex4}
\begin{document}

\title{Generalized Seesaw Mechanism of Neutrino and Bose-Einstein Condensation in the Modified O'Raifeartaigh Model}
\author{Tadafumi Ohsaku}
\affiliation{Institut f\"{u}r Theoretische Physik, Universit\"{a}t zu K\"{o}ln, 50937 K\"{o}ln, Germany}
\email[Present address: Department of Physics, University of Texas at Austin,]{tadafumi@physics.utexas.edu}
\date{\today}

\newcommand{\bmx}{\mbox{\boldmath $x$}}
\newcommand{\bmp}{\mbox{\boldmath $p$}}
\newcommand{\bmk}{\mbox{\boldmath $k$}}
\newcommand{\kfey}{\ooalign{\hfil/\hfil\crcr$k$}}
\newcommand{\pfey}{\ooalign{\hfil/\hfil\crcr$p$}}
\newcommand{\partfey}{\ooalign{\hfil/\hfil\crcr$\partial$}}
\newcommand{\hfey}{\ooalign{\hfil/\hfil\crcr$h$}}

\begin{abstract}

The modified O'Raifeartaigh model from the context of 
the generalized seesaw mechanism of neutrino mass is investigated.
In our evaluation of effective potentials of the theory, 
both the component field and the superspace formalisms to approach the problem are presented.
In the component field formalism, 
we take into account the Bose-Einstein condensates in the scalar sector by the method of many-boson theory, 
i.e. we consider both the condensates and the Hartree-Fock-Bogoliubov-type self-energies 
of quantum fluctuations.
The diagonalization of the mass matrix of the fermion sector gives 
the same functional forms of the mass eigenvalues in the generalized seesaw mechanism. 
The stability condition in the vicinity of the classical vacuum 
which shows the generalized seesaw situation
is obtained by the examination of the mass eigenvalues of the scalar sector of the model.
The superspace formalism will be devoted to a comparison between 
its result with that of the component field formalism.
(Keywords: Supersymmetric Effective Theories, Neutrino Physics, Supersymmetry Breaking, Nonperturbative Effects)

\end{abstract}


\maketitle

\section{Introduction}

The standard model of elementary particles is the most important achievement in modern physics,
and still it gives us the horizon of particle phenomenology~[1].
Supersymmetry ( SUSY )~[2-22] can be understood as one of the ways toward "beyond the standard model", 
from a viewpoint of particle phenomenology.
In such a supersymmetric approach, a theory has both fermionic and bosonic degrees of freedom, 
and they interact with each other under a supersymmetric manner.
Dynamics of interacting boson gases quite often show Bose-Einstein condensation ( BEC ) 
as a universal phenomenon. 
BEC was first found by Bose~[23] and Einstein~[24],
and theory of BEC of an interacting nonrelativistic boson gas was constructed firstly by Bogoliubov~[25].
The Bogoliubov theory has a very universal character, 
and it is the case that the theory can be applied to various interacting boson systems.
Moreover, methods and concepts of BEC and superfluidity of boson gases 
are useful to examine/understand an interacting fermion system,
for example, 
BCS ( Bardeen-Cooper-Schrieffer ) superconductivity~[26] or
chiral condensations in the Nambu$-$Jona-Lasinio model ( NJL )~[27,28] 
and quantum chromodynamics ( QCD )~[29], from the context of spontaneous symmetry breakings. 
On the other hand, a SUSY multiplet must be broken in a model for phenomenology
because we have not yet found any superpartner.
Due to the ( perturbative ) nonrenormalization theorem, 
a SUSY breaking cannot take place in a perturbation theory,
and it should be realized in a nonperturbative manner, i.e. a spontaneous SUSY breaking.
A lot of modern particle theoreticians consider that 
a dynamical symmetry breaking is phenomenologically prefered for a SUSY breakdown~[3,4,5,7,9,10,12].

\vspace{3mm}

The most important problem in modern particle physics which has been found by experimental results 
is the origin of masses, their hierarchy, and flavor violations of particles.
Recent experimental observations confirmed that neutrinos should have very tiny masses,
and the seesaw mechanism is one of candidates for providing an explanation to neutrino masses~[30-34].
Hence, it is an interesting issue to make a SUSY model which will show a seesaw machanism.
In the ordinary seesaw mechanism, neutrino has both a Dirac and a right-handed Majorana mass terms.
The references (35) and (36) discussed a generalization of the ordinary seesaw mechanism,
added a ( very tiny ) left-handed Majorana mass, and some interesting results were obtained.
It is well-known fact that the O'Raifeartaigh model breaks SUSY at its tree level~[14].
Recently, the modified O'Raifeartaigh model has been examined 
in the context of meta-stable SUSY breaking~[10-13].
Reference (11) gives mass eigenvalues of scalars and spinors: 
In fact the eigenvalues take quite similar structure
with that of the generalized seesaw mechanism~[35,36].
The purpose of this paper is to examine the modified O'Raifeataigh model 
under the context of the generalized seesaw mechanism of neutrinos.

\vspace{3mm}

This paper is organized as follows. 
In Sec. II, we invesitigate the generalized seesaw mechanism of the so-called modified O'Raifeartaigh model
in the component field formalism.
It is suitable to employ several many-body-theoretical techniques in the component field formalism 
though it becomes more lengthy than the superspace formalism.
While, we have to use the notion of component fields at some discussions also in the superspace formalism,
especially if we wish to take into account BEC in the scalar sector of the theory.
After introducing the modified O'Raifeartaigh model, 
we shortly discuss its symmetry property and the classical minimum.
We consider it might be possible that 
the solution of the classical minimum shows the generalized seesaw mass relation.
Then, we employ the many-body-theoretical technique to take into account BEC in the scalar sector.
By these preparation, the one-loop effective potential is calculated, 
and stability around the classical minimum will be investigated.
Possibility of SUSY breakdown around the classical minimum also be examined.
In a one-loop effective potential calculation, the loop expansion must converge rapidly enough,
and thus the vacuum of a theory should have a semiclassical nature, 
will not obtain a radical modification by possible quantum corrections.  
This must be the case in our calculation, 
and thus we consider the situation where quantum corrections around a classical minimum are small.
For comparison/supplement to the result of the component field formalism, 
a calculation of the one-loop effective potential in the superspace formalism is given in Sec. III.
The summary and conclusion of this work is presented in Sec. IV.

\vspace{3mm}

We will follow the textbook of Wess and Bagger for the spinor algebra, 
gamma matrices and metric conventions throughout this paper~[2].
( For example, the metric is $\eta^{\mu\nu}={\rm diag}(-1,1,1,1)$. )

\section{Component Field Formalism}

\subsection{The Classical Solution}

Our starting point is the following Lagrangian of the modified O'Raifeartaigh model~[10-13] 
of three chiral matter fields:
\begin{eqnarray}
{\cal L} &=& \Bigl( X^{\dagger}X + \Phi^{\dagger}_{+}\Phi_{+} + \Phi^{\dagger}_{-}\Phi_{-} \Bigr)\Big|_{\theta\theta\bar{\theta}\bar{\theta}}    \nonumber \\
& & + \Bigl( fX + \frac{g}{2}X\Phi_{+}\Phi_{+} + m_{D}\Phi_{+}\Phi_{-} + m_{L}\Phi_{-}\Phi_{-} \Bigr)\Big|_{\theta\theta}   \nonumber \\
& & + \Bigl( f^{\dagger}X^{\dagger} + \frac{g^{\dagger}}{2}X^{\dagger}\Phi^{\dagger}_{+}\Phi^{\dagger}_{+} + m^{\dagger}_{D}\Phi^{\dagger}_{+}\Phi^{\dagger}_{-} + m^{\dagger}_{L}\Phi^{\dagger}_{-}\Phi^{\dagger}_{-} \Bigr)\Big|_{\bar{\theta}\bar{\theta}}.
\end{eqnarray}
Here, $X$, $\Phi_{\pm}$ are chiral ( $+$; right, $-$; left ) superfields.
We regard $\Phi_{\pm}$ as neutrino superfields, 
$m_{D}$ and $m_{L}$ denote a Dirac and a left-handed Majorana mass parameters, respectively.
If $X$ takes a ( very large ) VEV compared with $m_{D}$ and $m_{L}$, 
then the theory may show a seesaw-type situation in
the mass matrix eigenvalues of its fermion sector.
The usual ( ordinary ) seesaw situation will be achieved by $m_{L}\to 0$.
The mass dimensions of $f$ and $g$ become as follows: $f;[{\rm mass}]^{2}$, $g;[{\rm mass}]^{0}$.
We consider the following global $U(1)_{V}$ ( gauge ) and $U(1)_{A}$ ( chiral ) transformations:
\begin{eqnarray}
& & U(1)_{V}: \, \Phi_{+} \to e^{i\alpha_{V}}\Phi_{+}, \quad \Phi_{-} \to e^{-i\alpha_{V}}\Phi_{-},  \quad
U(1)_{A}: \, \Phi_{+} \to e^{i\alpha_{A}}\Phi_{+}, \quad \Phi_{-} \to e^{i\alpha_{A}}\Phi_{-},  \quad
\alpha_{V}, \alpha_{A} \in {\bf R}.
\end{eqnarray}
The Majorana mass term of the mass parameter $m_{L}$ explicitly breaks both of these global symmetries.
We can choose the charge of $X$ to keep the coupling term $\frac{g}{2}X\Phi_{+}\Phi_{-}$ invariant under these transformations as
\begin{eqnarray}
U(1)_{V}: \, X \to e^{-2i\alpha_{V}}X, \quad U(1)_{A}: \, X \to e^{-2i\alpha_{A}}X.
\end{eqnarray}
The terms $fX$ and $f^{\dagger}X^{\dagger}$ also explicitly break these global $U(1)$ symmetries.
The Majorana mass term will breaks the $U(1)_{R}$ symmetry 
under the following charge assignment of the superfields $X$ and $\Phi_{\pm}$~[10-13]:
\begin{eqnarray}
U(1)_{R}: \, \theta \to e^{-i\alpha_{R}}\theta, \quad \bar{\theta} \to e^{i\alpha_{R}}\bar{\theta}, \quad 
X \to e^{2i\alpha_{R}}X, \quad \Phi_{+} \to \Phi_{+}, \quad \Phi_{-} \to e^{2i\alpha_{R}}\Phi_{-}, \quad
\alpha_{R} \in {\bf R}.
\end{eqnarray}
Another $R$-charge assignment is also possible:
\begin{eqnarray}
U(1)_{R}: \, X \to X, \quad \Phi_{+} \to e^{i\alpha_{R}}\Phi_{+}, \quad \Phi_{-} \to e^{i\alpha_{R}}\Phi_{-}.
\end{eqnarray}
In this case, the $R$-symmetry will be restored at the limit $f\to 0$.
Therefore, the term $fX$ and the Majorana mass term are incompatible with respect to the $U(1)_{R}$ symmetry.
As a result, there is no global $U(1)$ symmetry in our theory.
The absence of $R$-axion in the modified O'Raifeartaigh model is discussed in Refs.~[11,12] 
from the context of meta-stable SUSY breaking, and it is phenomenologically favorable.
The ordinary O'Raifeartaigh model corresponds to the case $m_{L}=m^{\dagger}_{L}=0$, it has an $R$-symmetry,
and SUSY is broken at the tree level~[12,14]. 
In the ordinary O'Raifeartaigh model, the classical solution becomes
$\phi_{+}=\phi_{-}=0$ with $\phi_{X}=$ arbitrary, and SUSY is spontaneously broken in the vacuum.
The one-loop effective potential of the O'Raifeartaigh model was calculated in Ref.~[15].
In that calculation, the degeneracy of vacua is lifted by the one-loop correction,
and the origin of the potential becomes the only ground state.
There is a ${\bf Z}_{2}$ symmetry under $\Phi_{\pm}\to -\Phi_{\pm}$ in (1).
The tree level part of the Lagrangian will be obtained from the scalar potential by employing the Euler-Lagrange equations of the auxiliary fields of chiral multiplets: 
\begin{eqnarray}
V^{tree}[\phi_{\pm},\phi_{X}] &=& |F_{X}|^{2} + |F_{+}|^{2} + |F_{-}|^{2} 
= \Big|f+\frac{g}{2}\phi^{2}_{+}\Big|^{2} + \Big|g\phi_{X}\phi_{+}+m_{D}\phi_{-}\Big|^{2} + \Big|m_{D}\phi_{+}+2m_{L}\phi_{-} \Big|^{2}.
\end{eqnarray}
In literature, the classical solution of $V^{tree}$ is given as follows~[??]
\begin{eqnarray}
\phi^{classical}_{X} &=& \frac{m^{2}_{D}}{2gm_{L}}, \quad \phi^{classical}_{+} = \pm\sqrt{\frac{-2f}{g}}, \quad \phi^{classical}_{-} = \mp\frac{m_{D}}{2m_{L}}\sqrt{\frac{-2f}{g}}.
\end{eqnarray}
Here we have assumed that $m_{D}$, $m_{L}$, $g$ and $f$ are real valued to obtain the classical minimum.
When these parameters are real, the classical solution shows the spontaneous ${\bf Z}_{2}$ symmetry breakdown.
All of the VEVs of the solution will go to infinity at the limit $g\to 0$, 
and $g=0$ is a singular point for the solution.
( This expression of the classical minimum of our model has a similarity with the classical solution of a Ginzburg-Landau-type $\varphi^{4}$ model
which describes the low-energy property of the Ising ferromagnet~[37]. 
Hence $V^{tree}$ seems to have a relation with the Ising ferromagnet. )
$V^{tree}$ vanishes at the classical minimum and 
the ${\cal N}=1$ SUSY of this model is unbroken at the classical level~[11,12].
We will see quantum corrections to the classical solution through the following loop expansion calculation,
though we mainly investigate a possibiliy of the generalized seesaw situation in the vicinity of the classical solution 
in this work.
As mentioned above, 
the ordinary O'Raifeartaigh model gives $\langle\phi_{X}\rangle$=arbitray at its classical solution
with broken SUSY, and the one-loop correction gives the unique vacuum as the origin of the potential
( the origin of (6) gives a finite energy and not supersymetric ).
By introducing the left-handed Majorana mass term in (1),
$\phi^{classical}_{X}$ has obtained the explicit expression given in (7) and then we can discuss
the strength of the VEV $\langle \phi_{X}\rangle$ which would give 
a right-handed Majorana mass parameter with some dynamics of our theory:
If we set $m_{L}=0$ from the beginning of our model, 
we cannot obtain an explicit expression for $\langle\phi_{X}\rangle$
( at least at the classical level ).
This is crucial in the context of this work.
The ordinary seesaw mechanism ( the case $m_{L}=0$ ) cannot be considered by the ordinary O'Raifeartaigh model. 
In this paper, we examine
(i) when the classical solution can give the generalized seesaw situation,
(ii) how the vicinity of the classical soution in the one-loop potential is stable and robust,
(iii) whether the vicinity of the classical solution in the one-loop potential breaks SUSY or not.  
Usually, SUSY is broken if there is an $R$-symmetry in a theory, 
while SUSY will be kept if an $R$-symmetry is broken.

\vspace{3mm}

After eliminating the auxiliary fields of $X$ and $\Phi_{\pm}$ and perform integrations of Grassmann coordinates, 
one finds the expression of ${\cal L}$ in terms of component fields as follows:
\begin{eqnarray}
{\cal L} &=& -\partial_{\nu}\phi^{\dagger}_{X}\partial^{\nu}\phi_{X} -\partial_{\nu}\phi^{\dagger}_{+}\partial^{\nu}\phi_{+} -\partial_{\nu}\phi^{\dagger}_{-}\partial^{\nu}\phi_{-}  
-i\bar{\psi}_{X}\bar{\sigma}^{\nu}\partial_{\nu}\psi_{X} -i\bar{\psi}_{+}\bar{\sigma}^{\nu}\partial_{\nu}\psi_{+} -i\bar{\psi}_{-}\bar{\sigma}^{\nu}\partial_{\nu}\psi_{-}   \nonumber \\
& & -|m_{D}|^{2}(|\phi_{+}|^{2}+|\phi_{-}|^{2}) -|g|^{2}|\phi_{X}|^{2}|\phi_{+}|^{2} - 4|m_{L}|^{2}|\phi_{-}|^{2} 
-\frac{|g|^{2}}{4}|\phi_{+}|^{4} -\frac{1}{2}(f^{\dagger}g\phi^{2}_{+}+fg^{\dagger}\phi^{\dagger 2}_{+}) -|f|^{2}  \nonumber \\
& & -(g^{\dagger}m_{D}\phi^{\dagger}_{X}+2m_{L}m^{\dagger}_{D})\phi^{\dagger}_{+}\phi_{-} -(gm^{\dagger}_{D}\phi_{X}+2m^{\dagger}_{L}m_{D})\phi^{\dagger}_{-}\phi_{+}  \nonumber \\
& & -\frac{g}{2}\phi_{X}\psi_{+}\psi_{+} -g\phi_{+}\psi_{X}\psi_{+} -m_{D}\psi_{+}\psi_{-} -m_{L}\psi_{-}\psi_{-} -\frac{g^{\dagger}}{2}\phi^{\dagger}_{X}\bar{\psi}_{+}\bar{\psi}_{+} -g^{\dagger}\phi^{\dagger}_{+}\bar{\psi}_{X}\bar{\psi}_{+} -m^{\dagger}_{D}\bar{\psi}_{+}\bar{\psi}_{-} -m^{\dagger}_{L}\bar{\psi}_{-}\bar{\psi}_{-}.
\end{eqnarray}
The phases of mass parameters and $\phi_{X}$ are defined as 
\begin{eqnarray}
\phi_{X} = |\phi_{X}|e^{i\theta_{X}}, \quad  m_{D} = |m_{D}|e^{i\theta_{D}}, \quad m_{L} = |m_{L}|e^{i\theta_{L}}, \quad
\theta_{X}, \theta_{D}, \theta_{L} \in {\bf R}. 
\end{eqnarray}
We can absorb only two of these phases $\theta_{X}$, $\theta_{D}$ and $\theta_{L}$ by a redefinition of fields.
Hereafter, we set $m_{D}=m^{\dagger}_{D}$ and $m_{L}=m^{\dagger}_{L}$ by a field redefinition 
while keeping the phase degree of freedom of $\phi_{X}$ without loss of generality. 
Later, we will observe that mass eigenvalues of scalars and spinors are functions of $\theta_{X}$. 
In principle, if we take into account phase degrees of freedom of $\phi_{\pm}$ and $\phi_{X}$,
the classical solution of $V^{tree}$ becomes 
( $f^{\dagger}=f$, $g^{\dagger}=g$, $m^{\dagger}_{D}=m_{D}$, $m^{\dagger}_{L}=m_{L}$ are imposed )
\begin{eqnarray}
|\phi_{+}| &=& \pm \sqrt{\frac{-2f}{g}}\sqrt{\cos(2\theta_{+})\pm\sqrt{\cos^{2}(2\theta_{+})-1}},   \nonumber \\
|\phi_{-}| &=& -\frac{m_{D}|\phi_{+}|}{2m_{L}}\Bigl[\cos(\theta_{+}-\theta_{-})\pm\sqrt{\cos^{2}(\theta_{+}-\theta_{-})-1}\Bigr],   \nonumber \\
|\phi_{X}| &=& -\frac{m_{D}|\phi_{+}|}{g|\phi_{+}|}\Bigl[\cos(\theta_{X}+\theta_{+}-\theta_{-})\pm\sqrt{\cos^{2}(\theta_{X}+\theta_{+}-\theta_{-})-1}\Bigr], \qquad  
\phi_{\pm} = |\phi_{\pm}|e^{i\theta_{\pm}}.  
\end{eqnarray}
However, the phase degrees of freedoms of scalars are chosen 
by the vanishing conditions of square roots in (10) such as 
$\theta_{+}=0,\pi$, $\theta_{-}=0,\pi$, $\theta_{X}=0,\pi$ ( totally 16 solutions, all are degenerate ).
The solutions (7) at the classical level are special cases of them.
On the contrary, later, 
we show it is important to take into account the phase degrees of freedom $\theta_{X}$,
by our examination of particle mass eigenvalues at one-loop level.
It is quite difficult to take into account phase degrees of freedom of scalar fields and mass parameters
in a complete manner in our calculation of the one-loop level ( and also, possible renormalization to them ), 
and thus we will use (7) frequently in our discussion.
Due to the Hermiticity of our Lagrangian, 
we have obtained the quartic terms $|g|^{2}|\phi_{+}|^{4}/4$ and $|g|^{2}|\phi_{+}|^{2}|\phi_{X}|^{2}$ 
as positive ( more precisely, non-negative ) definite in (8).
This fact guarantees the convergence of functional integral of variable $\phi_{+}$ in Euclidean region.
These quartic interactions are hard-core repulsive interactions at $|g|>0$, 
give a stability of the scalar sector.

\subsection{Bose-Einstein Condensation}

From the examination at the tree-level of our theory, 
we speculate that a BEC takes place 
in the scalar sector of the effective potential of (1) also in the one-loop level.
To take into account the BEC under an appropriate manner, 
the scalar fields will be divided into the condensates and their fluctuation parts:
\begin{eqnarray} 
\phi_{X} = \phi^{c}_{X} + \tilde{\phi}_{X}, \quad \phi_{+} = \phi^{c}_{+} + \tilde{\phi}_{+}, \quad \phi_{-} = \phi^{c}_{-} + \tilde{\phi}_{-}, 
\end{eqnarray}
where the superscript $c$ indicates the condensation parts of the fields.
We should mention that 
the classical solution and Bose-Einstein condensates 
of $\phi_{X}$ and $\phi_{\pm}$ are different in principle~[38],
\begin{eqnarray}
\phi^{classical}_{X} \ne \phi^{c}_{X}, \quad \phi^{classical}_{\pm} \ne \phi^{c}_{\pm},
\end{eqnarray}
because the latter include quantum corrections~[38].
We assume condensates are space-time independent.
Under the decomposition of (11), one finds
\begin{eqnarray}
& & |\phi_{\pm}|^{2} \to |\tilde{\phi}_{\pm}|^{2} + \phi^{c\dagger}_{\pm}\tilde{\phi}_{\pm} + \tilde{\phi}^{\dagger}_{\pm}\phi^{c}_{\pm} + |\phi^{c}_{\pm}|^{2}, \nonumber \\
& & \phi_{+}^{2} \to (\tilde{\phi}_{+})^{2} + 2\phi^{c}_{+}\tilde{\phi}_{+} + (\phi^{c}_{+})^{2},  \quad
\phi_{+}^{\dagger 2} \to (\tilde{\phi}^{\dagger}_{+})^{2} + 2\phi^{c\dagger}_{+}\tilde{\phi}^{\dagger}_{+} + (\phi^{c\dagger}_{+})^{2}, \quad \cdots, 
\end{eqnarray}
so forth. 
Consequently, for example, 
the Dirac and Majorana mass terms of the scalar sector give terms linear in $\tilde{\phi}_{\pm}$.
In fact, $m_{D}$ and $m_{L}$ have a role similar to chemical potential of a nonrelativistic boson theory.
The terms linear in the fluctuating fields ( and tadpole-type diagrams ) will be dropped from our Lagrangian.
This "variational" condition corresponds to the Euler-Lagrange equations for condensates~[37-42]. 
The quartic interactions of scalars in ${\cal L}$ become
\begin{eqnarray}
\frac{|g|^{2}}{4}|\phi_{+}|^{4} &=& \frac{|g|^{2}}{4}\Bigl[ |\tilde{\phi}_{+}|^{4} + 4|\tilde{\phi}_{+}|^{2}|\phi^{c}_{+}|^{2} + |\phi^{c}_{+}|^{4}
+ 2 |\tilde{\phi}_{+}|^{2} (\tilde{\phi}_{+}\phi^{c\dagger}_{+} + \tilde{\phi}^{\dagger}_{+}\phi^{c}_{+}) 
+ \tilde{\phi}_{+}\tilde{\phi}_{+}\phi^{c\dagger}_{+}\phi^{c\dagger}_{+}
+ \tilde{\phi}^{\dagger}_{+}\tilde{\phi}^{\dagger}_{+}\phi^{c}_{+}\phi^{c}_{+} \Bigr],  \nonumber \\
|g|^{2}|\phi_{X}|^{2}|\phi_{+}|^{2} &=& |g|^{2}\Bigl[ |\phi^{c}_{X}|^{2}|\phi^{c}_{+}|^{2} + |\phi^{c}_{X}|^{2}|\tilde{\phi}_{+}|^{2} 
+ |\phi^{c}_{+}|^{2}|\tilde{\phi}_{X}|^{2} + |\tilde{\phi}_{X}|^{2}|\tilde{\phi}_{+}|^{2} 
+ \phi^{c\dagger}_{X}\phi^{c}_{+}\tilde{\phi}^{\dagger}_{+}\tilde{\phi}_{X}
+ \phi^{c\dagger}_{+}\phi^{c}_{X}\tilde{\phi}^{\dagger}_{X}\tilde{\phi}_{+}   \nonumber \\
& & + \phi^{c\dagger}_{X}\phi^{c\dagger}_{+}\tilde{\phi}_{+}\tilde{\phi}_{X}
+ \phi^{c}_{+}\phi^{c}_{X}\tilde{\phi}^{\dagger}_{X}\tilde{\phi}^{\dagger}_{+} 
+ \phi^{c\dagger}_{X}|\tilde{\phi}_{+}|^{2}\tilde{\phi}_{X} + \phi^{c}_{X}|\tilde{\phi}_{+}|^{2}\tilde{\phi}^{\dagger}_{X}
+ \phi^{c\dagger}_{+}|\tilde{\phi}_{X}|^{2}\tilde{\phi}_{+} + \phi^{c}_{+}|\tilde{\phi}_{X}|^{2}\tilde{\phi}^{\dagger}_{+} \Bigr].
\end{eqnarray}
Here, we have dropped the terms linear in $\tilde{\phi}_{+}$ 
or $\tilde{\phi}_{X}$ and their Hermitian conjugates.
We will employ the Hartree-Fock-Bogoliubov ( HFB ) approximation to self-energies coming from
the quartic and cubic interactions between fluctuations with introducing the following vacuum expectation values:
\begin{eqnarray}
& & J_{1}(x) \equiv \langle\tilde{\phi}^{\dagger}_{+}(x)\tilde{\phi}_{+}(x) \rangle, \quad 
J_{2}(x) \equiv \langle\tilde{\phi}_{+}(x)\tilde{\phi}_{+}(x) \rangle, \quad
J^{\dagger}_{2}(x) \equiv \langle\tilde{\phi}^{\dagger}_{+}(x)\tilde{\phi}^{\dagger}_{+}(x) \rangle,  \quad
K_{1}(x) \equiv \langle\tilde{\phi}^{\dagger}_{X}(x)\tilde{\phi}_{X}(x) \rangle, \nonumber \\
& & K_{2}(x) \equiv \langle\tilde{\phi}_{X}(x)\tilde{\phi}_{+}(x) \rangle, \quad
K^{\dagger}_{2}(x) \equiv \langle\tilde{\phi}^{\dagger}_{+}(x)\tilde{\phi}^{\dagger}_{X}(x) \rangle,  \quad
K_{3}(x) \equiv \langle\tilde{\phi}^{\dagger}_{X}(x)\tilde{\phi}_{+}(x) \rangle, \quad 
K_{3}^{\dagger}(x) \equiv \langle\tilde{\phi}^{\dagger}_{+}(x)\tilde{\phi}_{X}(x) \rangle.
\end{eqnarray}
Here, $J_{1}=\langle \tilde{\phi}^{\dagger}_{+}\tilde{\phi}_{+} \rangle$ is normal, while
$J_{2}=\langle \tilde{\phi}_{+}\tilde{\phi}_{+}\rangle$ and $J^{\dagger}_{2} = \langle \tilde{\phi}^{\dagger}_{+}\tilde{\phi}^{\dagger}_{+}\rangle$ are anomalous self-energies, 
similar notions to the case of the BCS-Nambu-Gor'kov theory of superconductivity~[26,43,44].
The anomalous self-energies indicate a breakdown of particle-number non-conservation in the scalar sector,
and this is of course an independent phenomenon with 
the particle-number-non-conservation caused by the Majorana mass term of the fermion sector. 
In nonrelativistic theory of BEC, anomalous self-energies are negative quantities.
Therefore, one obtaines
\begin{eqnarray}
|\tilde{\phi}_{+}|^{2}\tilde{\phi}_{+} &\to& 2J_{1} \tilde{\phi}_{+} + J_{2} \tilde{\phi}^{\dagger}_{+},   \quad
|\tilde{\phi}_{+}|^{2}\tilde{\phi}^{\dagger}_{+} \to 2J_{1} \tilde{\phi}^{\dagger}_{+} + J^{\dagger}_{2} \tilde{\phi}_{+},  \quad
|\tilde{\phi}_{+}|^{4} \to 4 J_{1} \tilde{\phi}^{\dagger}_{+}\tilde{\phi}_{+} + J^{\dagger}_{2} \tilde{\phi}_{+}\tilde{\phi}_{+} + J_{2} \tilde{\phi}^{\dagger}_{+}\tilde{\phi}^{\dagger}_{+}.
\end{eqnarray}
By the HFB approximation, the cubic interactions of fluctuations will also be dropped from our Lagrangian. 
Hence we get
\begin{eqnarray}
\frac{|g|^{2}}{4}|\phi_{+}|^{4} &\to& \frac{|g|^{2}}{4}\Bigg[  
\bigl( 4J_{1} + 4|\phi^{c}_{+}|^{2} \bigr)\tilde{\phi}^{\dagger}_{+}\tilde{\phi}_{+}
+ \bigl( J^{\dagger}_{2} + (\phi^{c\dagger}_{+})^{2} \bigr)\tilde{\phi}_{+}\tilde{\phi}_{+}
+ \bigl( J_{2} + (\phi^{c}_{+})^{2} \bigr)\tilde{\phi}^{\dagger}_{+}\tilde{\phi}^{\dagger}_{+} + |\phi^{c}_{+}|^{4} \Bigg],
\end{eqnarray}
and
\begin{eqnarray}
|g|^{2}|\phi_{X}|^{2}|\phi_{+}|^{2} &\to& |g|^{2}\Bigg[ |\phi^{c}_{X}|^{2}|\phi^{c}_{+}|^{2} + ( |\phi^{c}_{X}|^{2} + K_{1} )|\tilde{\phi}_{+}|^{2} 
+ ( |\phi^{c}_{+}|^{2} + J_{1} )|\tilde{\phi}_{X}|^{2}   \nonumber \\
& & 
+ ( \phi^{c\dagger}_{X}\phi^{c}_{+} + K_{3}  ) \tilde{\phi}^{\dagger}_{+}\tilde{\phi}_{X}
+ ( \phi^{c\dagger}_{+}\phi^{c}_{X} + K^{\dagger}_{3}  ) \tilde{\phi}^{\dagger}_{X}\tilde{\phi}_{+}
+ ( \phi^{c\dagger}_{X}\phi^{c\dagger}_{+} + K^{\dagger}_{2} ) \tilde{\phi}_{+}\tilde{\phi}_{X}
+ ( \phi^{c}_{+}\phi^{c}_{X} + K_{2}  ) \tilde{\phi}^{\dagger}_{X}\tilde{\phi}^{\dagger}_{+} \Bigg].
\end{eqnarray}
From the classical solution,
we guess $|g|^{2}|\phi^{c}_{+}|^{2}$ and $|g|^{2}|\phi^{c}_{X}|^{2}$ take values of 
$\sim{\cal O}(|g|^{0})$, 
and thus a matrix elements given by a polynomial of them are not small enough to neglect from our Lagrangian.
While, we hope $0 < |g| \ll 1$ to be satisfied 
for convergence of a perturbative series/diagrams in terms of $|g|$.
This condition could conflict with the seesaw condition
$g\langle \phi_{X}\rangle\gg m_{D}\gg m_{L}$:
A realization of the generalized seesaw situation is a non-trivial problem in our theory.
In the next subsection, we will evaluate the one-loop effective potential of our theory.
We hope the potential captures the essential feature of quantum dynamics of the system (1) even at the one-loop level.
On the other hand, we simply drop $-\frac{g}{2}(\tilde{\phi}_{X}\psi_{+}\psi_{+}+\tilde{\phi}_{+}\psi_{X}\psi_{+})+({\rm h.c.})$ 
which will give a coupling between the scalar and spinor sectors in ${\cal L}$.

\vspace{3mm}

Now, we examine the variational condition, 
namely the vanishing condition of the terms linear in fluctuating scalars.
From (8), we obtain the linear terms as follows:
\begin{eqnarray}
& & -\tilde{\phi}_{+}
\Bigl[ 
|m_{D}|^{2}\phi^{c\dagger}_{+} + f^{\dagger}g\phi^{c}_{+}
+ gm^{\dagger}_{D}\phi^{c}_{X}\phi^{c\dagger}_{-} + 2m^{\dagger}_{L}m_{D}\phi^{c\dagger}_{-}  \nonumber \\
& & \qquad  + |g|^{2}
\Bigl( 
\frac{ |\phi^{c}_{+}|^{2} + J_{1} + 2|\phi^{c}_{X}|^{2} + 2K_{1} }{2}\phi^{c\dagger}_{+} 
+ \frac{J^{\dagger}_{2} + 2K^{\dagger}_{3} }{2}\phi^{c}_{+} 
+ K^{\dagger}_{2}\phi^{c}_{X} + K^{\dagger}_{3}\phi^{c\dagger}_{X} 
\Bigr)
\Bigr]   \nonumber \\
& & -\tilde{\phi}_{-}
\Bigl[ 
(|m_{D}|^{2}+4|m_{L}|^{2})\phi^{c\dagger}_{-}
+ (g^{\dagger}m_{D}\phi^{c\dagger}_{X}+2m_{L}m^{\dagger}_{D})\phi^{c\dagger}_{+} 
\Bigr]   \nonumber \\
& & -\tilde{\phi}_{X}
\Bigl[
gm^{\dagger}_{D}\phi^{c}_{+}\phi^{c\dagger}_{-}+ |g|^{2}
\Bigl( 
J_{1}\phi^{c\dagger}_{X} + K^{\dagger}_{2}\phi^{c}_{+} 
\Bigr)
\Bigr]  \nonumber \\
& & + {\rm h.c.}
\end{eqnarray}
They must vanish in our treatment of BEC.
All terms given above have mass dimension $[Mass]^{4}$.
We will employ a kind of Popov approximation to our HFB theory~[38], i.e. $J_{2}=K_{2}=K_{3}=0$
( all of the anomalous self-energies will be dropped ).
At the classical solution (7), the vanishing condition of the coefficient function of $\tilde{\phi}_{X}$
gives
\begin{eqnarray} 
J_{1} &=& J^{\dagger}_{1} = -\frac{2f}{g}.
\end{eqnarray}
This expression would be modified under a renormalization of bare parameters as
$J_{1}=-2f^{(ren)}/g^{(ren)}$.
We confirm that the coefficient of $\tilde{\phi}_{-}$ vanishes identically at the classical solution (7).
From the vanishing condition of the coefficient of $\tilde{\phi}_{+}$ at (7), one finds
\begin{eqnarray}
K_{1} &=& K^{\dagger}_{1} = g^{-2}\Bigl( \frac{m^{4}_{D}+2m_{L}m^{3}_{D}}{4m^{2}_{L}} -gf \Bigr).
\end{eqnarray}
This expression of $K_{1}$ takes a large value at the seesaw condition $m_{D}\gg m_{L}$.

\vspace{3mm}

Next, we will introduce several fields of the following definitions for the convenience of our discussion:
\begin{eqnarray}
& & 
\Psi \equiv ( \Psi_{X},\Psi_{M} )^{T}, \quad  
\Psi_{X} \equiv (\psi_{X},\bar{\psi}_{X})^{T}, \quad 
\Psi_{M} \equiv (\Psi_{MR},\Psi_{ML})^{T}, \quad
\Psi_{MR} \equiv (\psi_{+},\bar{\psi}_{+})^{T}, \quad 
\Psi_{ML} \equiv (\psi_{-},\bar{\psi}_{-})^{T}, 
\nonumber \\
& & 
\overline{\Psi} \equiv ( \overline{\Psi}_{X}, \overline{\Psi}_{M} ), \quad 
\overline{\Psi}_{X} \equiv (-\psi_{X},-\bar{\psi}_{X}), \quad 
\overline{\Psi}_{M} \equiv (\overline{\Psi}_{MR},\overline{\Psi}_{ML}), \quad 
\overline{\Psi}_{MR} \equiv (-\psi_{+},-\bar{\psi}_{+}), \quad 
\overline{\Psi}_{ML} \equiv (-\psi_{-},-\bar{\psi}_{-}), 
\nonumber \\
& & 
\Pi \equiv (\Pi_{X},\Pi_{M})^{T}, \quad 
\Pi_{X} \equiv (\tilde{\phi}_{X},\tilde{\phi}^{\dagger}_{X})^{T}, \quad 
\Pi_{M} \equiv (\Pi_{MR},\Pi_{ML})^{T}, \quad 
\Pi_{MR} \equiv (\tilde{\phi}_{+},\tilde{\phi}^{\dagger}_{+})^{T}, \quad
\Pi_{ML} \equiv (\tilde{\phi}_{-},\tilde{\phi}^{\dagger}_{-})^{T}.
\end{eqnarray}
Here, $\psi_{X}$ is a Majorana, $\psi_{MR}$ and $\psi_{ML}$ are right- and left- handed Majorana fields, respectively. 
$T$ denotes transposition operation of a matrix.
The Lagrangian density will be rewritten in the following form by these fields:
\begin{eqnarray}
{\cal L} &=& -V^{tree}[\phi^{c}_{\pm},\phi^{c}_{X}] + \frac{1}{2}\Pi^{\dagger}\Omega^{B}\Pi + \frac{1}{2}\overline{\Psi}\Omega^{F}\Psi.
\end{eqnarray}
The matrices $\Omega^{B}$ and $\Omega^{F}$ are defined as follows:
\begin{eqnarray}
& & \Omega^{B} \equiv \left(
\begin{array}{cc}
\Omega^{B}_{XX} & \Omega^{B}_{XM}  \\
\Omega^{B}_{MX} & \Omega^{B}_{MM}
\end{array}
\right), \quad 
\Omega^{B}_{MM} \equiv \left(
\begin{array}{cc}
\Omega^{B}_{++} & \Omega^{B}_{+-} \\
\Omega^{B}_{-+} & \Omega^{B}_{--} 
\end{array}
\right), \quad 
\Omega^{B}_{XM} \equiv \bigl(
\Omega^{B}_{X+},\Omega^{B}_{X-} \bigr), \quad
\Omega^{B}_{MX} \equiv \left(
\begin{array}{c}
\Omega^{B}_{+X} \\
\Omega^{B}_{-X} 
\end{array}
\right),
\nonumber \\ 
& & \Omega^{F} \equiv \left(
\begin{array}{cc}
\Omega^{F}_{XX} & \Omega^{F}_{XM}  \\
\Omega^{F}_{MX} & \Omega^{F}_{MM}
\end{array}
\right), \quad \Omega^{F}_{XX} \equiv i\partfey, \quad \Omega^{F}_{XM} \equiv 
\bigl( -g\phi^{c}_{+}P_{+} -g^{\dagger}\phi^{c\dagger}_{+}P_{-}, 0 \bigr), \nonumber \\
& & \Omega^{F}_{MX} \equiv \left(
\begin{array}{c}
-g\phi^{c}_{+}P_{+} -g^{\dagger}\phi^{c\dagger}_{+}P_{-} \\
0 
\end{array}
\right), \quad \Omega^{F}_{MM} \equiv \left( 
\begin{array}{cc}
i\partfey -g\phi^{c}_{X}P_{+} -g^{\dagger}\phi^{c\dagger}_{X}P_{-} & -m_{D}  \\
-m_{D} & i\partfey -2m_{L} 
\end{array}
\right).
\end{eqnarray}
The definitions
$\gamma^{5} \equiv \gamma^{0}\gamma^{1}\gamma^{2}\gamma^{3}$ and  
$P_{\pm} \equiv \frac{1\pm i\gamma^{5}}{2}$ have been used.
The Hermiticity $(\Omega^{B})^{\dagger}=\Omega^{B}$ is satisfied. 
Entries of $\Omega^{B}$ become such that,
\begin{eqnarray}
& & \Omega^{B}_{XX} \equiv \left(
\begin{array}{cc}
\Box - |g|^{2}( J_{1} + |\phi^{c}_{+}|^{2} ) & 0 \\
0 & \Box - |g|^{2}( J_{1} + |\phi^{c}_{+}|^{2} )
\end{array}
\right), \quad \Omega^{B}_{--} \equiv \left(
\begin{array}{cc}
\Box - m^{2}_{D} -4m^{2}_{L}  &  0 \\
0  & \Box - m^{2}_{D} -4m^{2}_{L} 
\end{array}
\right), \nonumber \\ 
& & \Omega^{B}_{++} \equiv \left(
\begin{array}{cc}
\Box - m^{2}_{D} -|g|^{2}(J_{1}+K_{1}+ |\phi^{c}_{+}|^{2} + |\phi^{c}_{X}|^{2})  & -\frac{|g|^{2}}{4}( J_{2} + (\phi^{c}_{+})^{2})-\frac{fg^{\dagger}}{2} \\
-\frac{|g|^{2}}{4}( J^{\dagger}_{2} + (\phi^{c\dagger}_{+})^{2})-\frac{f^{\dagger}g}{2} & \Box - m^{2}_{D} -|g|^{2}(J_{1}+K_{1}+ |\phi^{c}_{+}|^{2} + |\phi^{c}_{X}|^{2}) 
\end{array}
\right), \nonumber \\
& & \Omega^{B}_{X+} \equiv \left(
\begin{array}{cc}
-|g|^{2}( K^{\dagger}_{3} + \phi^{c\dagger}_{+}\phi^{c}_{X}   )  & -g^{\dagger}m_{D}\phi^{c}_{-} - |g|^{2}( K_{2} + \phi^{c}_{+}\phi^{c}_{X} ) \\
-gm_{D}\phi^{c\dagger}_{-} - |g|^{2}( K^{\dagger}_{2} + \phi^{c\dagger}_{+}\phi^{c\dagger}_{X} ) &  -|g|^{2}( K_{3} + \phi^{c\dagger}_{X}\phi^{c}_{+}  )
\end{array}
\right), \nonumber \\
& & \Omega^{B}_{X-} \equiv \left(
\begin{array}{cc}
-g^{\dagger}m_{D}\phi^{c\dagger}_{+} & 0 \\
0 & -gm_{D}\phi^{c}_{+}
\end{array}
\right), \quad 
\Omega^{B}_{+-} \equiv \left(
\begin{array}{cc}
-m_{D}(g^{\dagger}\phi^{c\dagger}_{X}+2m_{L}) & 0 \\
0 & -m_{D}(g\phi^{c}_{X}+2m_{L})
\end{array}
\right).
\end{eqnarray}
All of the off-diagonal elements of $\Omega^{B}$ are coming from particle-number-non-conserving interactions and/or mean-fields
of ${\cal L}$ under the HFB approximation.
The diagonalizations of $\Omega^{B}$ and $\Omega^{F}$ will give "quasiparticle" excitation energy spectra 
of scalar and spinor fields in terms of the bare parameters/fields.
Especially we have an interest on whether the spinor $\psi_{X}$ becomes massive or not 
under the one-loop quantum correction. 
If there is no massless fermion, then there is no Nambu-Goldstone ( NG ) fermion, 
and the Nambu-Goldstone theorem implies the absence of spontaneous SUSY breaking in our theory~[9,12].

\subsection{The One-loop Effective Potential}

In this subsection, we will evaluate and examine the one-loop effective potential of our theory.
We obtain the generating functional of our theory as follows:
\begin{eqnarray}
{\cal Z} &\equiv& \int 
{\cal D}\tilde{\phi}_{X}{\cal D}\tilde{\phi}^{\dagger}_{X}{\cal D}\tilde{\phi}_{+}{\cal D}\tilde{\phi}^{\dagger}_{+}{\cal D}\tilde{\phi}_{-}{\cal D}\tilde{\phi}^{\dagger}_{-}
{\cal D}\psi_{X}{\cal D}\bar{\psi}_{X}{\cal D}\psi_{+}{\cal D}\bar{\psi}_{+}{\cal D}\psi_{-}{\cal D}\bar{\psi}_{-}    \nonumber \\
& & \times \exp\Bigg[i\int d^{4}x \Bigl\{ 
-V^{tree}[\phi^{c}_{\pm},\phi^{c}_{X}] +  \frac{1}{2}\Pi^{\dagger}\Omega^{B}\Pi + \frac{1}{2}\overline{\Psi}\Omega^{F}\Psi \Bigr\} + ({\rm source})\Bigg].
\end{eqnarray}
The one-loop contribution to the effective potential is evaluated to be
\begin{eqnarray}
& & V^{(1)}[\phi^{c}_{\pm},\phi^{c}_{X}] 
= V^{B(1)} + V^{F(1)}, \nonumber \\
& & V^{B(1)} \equiv \frac{i}{2}\ln{\rm Det}\Omega^{B} = \frac{i}{2}\ln{\rm Det}\Omega^{B}_{MM} + \frac{i}{2}\ln{\rm Det}\Bigl(\Omega^{B}_{XX}-\Omega^{B}_{XM}\frac{1}{\Omega^{B}_{MM}}\Omega^{B}_{MX}\Bigr), \nonumber \\
& & V^{F(1)} \equiv -\frac{i}{2}\ln{\rm Det}\Omega^{F} = -\frac{i}{2}\ln{\rm Det}\Omega^{F}_{MM} -\frac{i}{2}\ln{\rm Det}\Bigl(\Omega^{F}_{XX}-\Omega^{F}_{XM}\frac{1}{\Omega^{F}_{MM}}\Omega^{F}_{MX}\Bigr).
\end{eqnarray}
The effective action is found to be
\begin{eqnarray}
\Gamma_{(compo)} &\equiv& -i\ln{\cal Z} = \int d^{4}x\Bigl(-V^{tree}[\phi^{c}_{\pm},\phi^{c}_{X}] -V^{(1)}[\phi^{c}_{\pm},\phi^{c}_{X}] \Bigr).
\end{eqnarray}
If we perform the path integration of only $\Pi_{M}$ and $\Psi_{M}$, the generating functional becomes
\begin{eqnarray}
{\cal Z} &=& \int 
{\cal D}\tilde{\phi}_{X}{\cal D}\tilde{\phi}^{\dagger}_{X}{\cal D}\psi_{X}{\cal D}\bar{\psi}_{X}\exp\Bigg[i\int d^{4}x \Bigl\{ 
-V^{tree}[\phi^{c}_{\pm},\phi^{c}_{X}] - V^{(1)}_{M}  \nonumber \\
& & + \frac{1}{2}\Pi^{\dagger}_{X}\Bigl(\Omega^{B}_{XX}-\Omega^{B}_{XM}\frac{1}{\Omega^{B}_{MM}}\Omega^{B}_{MX}\Bigr)\Pi_{X} + \frac{1}{2}\overline{\Psi}_{X}\Bigl(\Omega^{F}_{XX}-\Omega^{F}_{XM}\frac{1}{\Omega^{F}_{MM}}\Omega^{F}_{MX}\Bigr)\Psi_{X} \Bigr\} \Bigg],
\end{eqnarray}
where,
\begin{eqnarray}
V^{(1)}_{M} \equiv V^{B(1)}_{M} + V^{F(1)}_{M}, \quad
V^{B(1)}_{M} \equiv \frac{i}{2}\ln{\rm Det}\Omega^{B}_{MM}, \quad
V^{F(1)}_{M} \equiv -\frac{i}{2}{\rm Det}\Omega^{F}_{MM}.
\end{eqnarray}
To obtain this expression of ${\cal Z}$, 
we can regard $\Pi_{X}$ and $\Psi_{X}$ as Grassmann-even and Grassmann-odd source fields 
in the Gaussian integrations of $\Pi_{M}$ and $\Psi_{M}$, respectively.

\vspace{3mm}

Let us examine the matrix $\Omega^{F}$.
Since $\Omega^{F}_{XX}$ is the inverse of propagator of massless fermion,
a perturbative expansion in terms of $(\Omega^{F}_{XX})^{-1}$ for handling ${\rm Tr}\ln\Omega^{F}$ 
suffers from infrared divergences, 
indicates that the perturbative expansion is an unsuitable method for our model, and thus it is forbidden.
If fermion $\psi_{X}$ remains massless at the one-loop level, 
the determinant of $\Omega^{F}$ has a zero point at $p^{2}=0$, 
must be factorized like $\det\Omega^{F}(p)=(p^{2})^{2}(p^{2}+({\rm mass})^{2})^{4}$. 
However, the direct evaluation of $\det\Omega^{F}(p_{\nu}=0)$ from (24), 
i.e. at the vanishing four-momentum, gives
\begin{eqnarray}
\det\Omega^{F}(p_{\nu}=0) &=& \bigl( 4|g|^{4}|\phi^{c}_{+}|^{4}m^{2}_{L} \bigr)^{2}.
\end{eqnarray} 
Hence, there is no massless particle in the fermion sector at $\phi^{c}_{+}\ne 0$,
and this fact indicates the absence of a spontaneous SUSY breakdown in our theory.
Note that this fact is globally the case, whole of the functional space of $\det\Omega^{F}$. 
The one-loop effective potential of the contribution of $\psi_{\pm}$ 
will be obtained after the diagonalization of $\Omega^{F}_{MM}$ in the following form:
\begin{eqnarray}
V^{F(1)}_{M} = -i{\rm Tr}\ln(k_{0}-E^{F}_{+})^{2}(k_{0}+E^{F}_{+})^{2}(k_{0}-E^{F}_{-})^{2}(k_{0}+E^{F}_{-})^{2}.
\end{eqnarray}
The energy spectra $E^{F}_{\pm}$ become
\begin{eqnarray}
E^{F}_{\pm}(\bmk) &\equiv& \sqrt{ \bmk^{2} + M^{F2}_{\pm}}, \nonumber \\
M^{F}_{\pm} &\equiv& \sqrt{ m_{D}^{2} + \frac{|g|^{2}|\phi^{c}_{X}|^{2}}{2} + 2m_{L}^{2} \mp 2 \sqrt{ \Bigl( \frac{|g|^{2}|\phi^{c}_{X}|^{2}}{4}-m_{L}^{2}\Bigr)^{2} + m_{D}^{2}\Bigl(\frac{|g|^{2}|\phi^{c}_{X}|^{2}}{4}+m_{L}^{2}+|g||\phi^{c}_{X}|m_{L}\cos\theta_{X}\Bigr) } }.
\end{eqnarray}
Here, the phase of $\phi_{X}$ appears in $M^{F}_{\pm}$ given above.
It was shown in Ref.~[36] that a one-loop potential is not degenerate with the phase $\theta_{X}$
in a Nambu$-$Jona-Lasinio-type dynamical model of the generalized seesaw mechanism.
At the case $|g|^{2}|\phi^{c}_{X}|^{2}\gg m^{2}_{D}\gg m^{2}_{L}$
( satisfied under $m_{D}\gg m_{L}$ in (7) ), 
these mass spectra show the generalized seesaw mechanism~[35,36],
$M^{F}_{+}$ is light while $M^{F}_{-}$ is heavy. At $m_{D}\gg m_{L}$, they become
\begin{eqnarray}
M^{F}_{+} \sim \sqrt{2}m_{L}, \quad M^{F}_{-} \sim \sqrt{2m^{2}_{D}+2m^{2}_{L}+|g|^{2}|\phi^{c}_{X}|^{2}}.
\end{eqnarray}
Hence, in the generalized seesaw mechanism, 
the light spinor aquires its mass of ${\cal O}(m_{L})$ while $m_{D}$ and $|\phi_{X}|$ have 
quite minor contributions to it.
By taking into account the result (31), 
we obtain the following mass formula of the fermion sector in terms of the bare quantities:
\begin{eqnarray}
\bigl( (M^{F}_{X})^{2}(M^{F}_{+})^{2}(M^{F}_{-})^{2} \bigr)^{2} &=& \bigl( 4|g|^{4}|\phi^{c}_{+}|^{4}m^{2}_{L} \bigr)^{2}.
\end{eqnarray}
The right hand side becomes $16f^{4}m^{4}_{L}$ at the classical solution, 
can take a small value at $f\to 0$ or $m_{L}\to 0$.
$m_{L}=0$ is the case of ordinary O'Raifeartaigh model,
and in that case $M^{F}_{X}=0$ takes place, indicates a breakdown of SUSY.
Hence, the expression of mass of $\psi_{X}$-field is found to be
\begin{eqnarray}
M^{F}_{X} &=& \frac{2|g|^{2}|\phi^{c}_{+}|^{2}m_{L}}{\sqrt{m^{4}_{D}+4|g|^{2}|\phi^{c}_{X}|^{2}m^{2}_{L}-4|g||\phi^{c}_{X}|m^{2}_{D}m_{L}\cos\theta_{X}}}.
\end{eqnarray}
Note that $M^{F}_{X}$ can become imaginary 
when inside the square root of the denominator of (36) takes a negative value. 
By putting the classical solution (7), one obtaines
\begin{eqnarray}
M^{F}_{X} &=& \frac{-4f|g|m_{L}}{m^{2}_{D}\sqrt{2(1-\cos\theta_{X})}}.
\end{eqnarray}
$M^{F}_{X}$ becomes very small and will behave as a pseudo-NG fermion when $f,|g|,m_{L}\ll m_{D}$,
and it vanishes at $m_{L}=0$, 
while $(M^{F}_{X})^{2}$ is always a potitive quantity.
It is an interesting fact that $M^{F}_{X}$ will diverge under $\theta_{X}\to 0$,
namely, not well-defined in the limit.
( In the case of a non-SUSY dynamical model of the generalized seesaw mechanism, 
$\theta_{X}=\pi$ is chosen as the vacuum state~[36]. )
Therefore, a careful examination on $\theta_{X}$ is important ( crucial ) in our theory.
As examined in the subsection A, $\theta_{X}$ of the classical solutions will take $0$ or $\pi$.
$M^{F}_{X}\sim{\cal O}(fgm_{L}/m^{2}_{D})$ if we choose $\theta_{X}=\pi$.
We assume there is no spontaneous Lorentz symmetry breaking in the fermion sector. 
As a result, the spectrum of $\psi_{X}$-particle must take the following Lorentz symmetric form:
\begin{eqnarray}
E^{F}_{X} &=& \sqrt{\bmk^{2}+M^{F2}_{X}}.
\end{eqnarray}
Therefore, we get the one-loop contribution of the fermion sector as
\begin{eqnarray}
V^{F(1)} = -\frac{i}{2}{\rm Tr}\ln(k_{0}-E^{F}_{X})^{2}(k_{0}+E^{F}_{X})^{2}(k_{0}-E^{F}_{+})^{2}(k_{0}+E^{F}_{+})^{2}(k_{0}-E^{F}_{-})^{2}(k_{0}+E^{F}_{-})^{2}.
\end{eqnarray}
We summarize the result of our analysis of the fermion sector:
(i) Our model at $m_{L}=0$ corresponds to the ordinary O'Raifeartaigh model, 
has $R$-symmetry and SUSY is broken at the ground state, and we have confirmed that an NG fermion appears~[9,12].
(ii) Our model at $m_{L}\ne 0$, 
namely the modified O'Raifeartaigh model~[11,12], 
will give the generalized seesaw mechanism at the point (7),
while $\psi_{X}$ has a finite mass and SUSY seems not broken.
We will proceed our examination to the boson sector of our theory.

\vspace{3mm}

In our treatment for the scalar sector, first we solve the secular equation $\det\Omega^{B}_{MM}=0$.
Though the secular equation is quartic in the d'Alembertian $\Box$, 
fortunately, we can diagonalize $\Omega^{B}_{MM}$ analytically 
because its secular equation will be factorized into a product of two quadratic equations of $\Box$.
The results is
\begin{eqnarray}
\det\Omega^{B}_{MM} &=& (k_{0}-E^{B}_{M1+})(k_{0}+E^{B}_{M1+})(k_{0}-E^{B}_{M1-})(k_{0}+E^{B}_{M1-})    \nonumber \\
& & \times (k_{0}-E^{B}_{M2+})(k_{0}+E^{B}_{M2+})(k_{0}-E^{B}_{M2-})(k_{0}+E^{B}_{M2-}).
\end{eqnarray}
There is no degeneracy in the spectra obtained from $\det\Omega^{B}_{MM}=0$.
Here, the energy eigenvalues become 
\begin{eqnarray}
E^{B}_{M1\pm}(\bmk) &\equiv& \sqrt{ \bmk^{2} + (M^{B}_{M1\pm})^{2}},  \quad E^{B}_{M2\pm}(\bmk) \equiv \sqrt{ \bmk^{2} + (M^{B}_{M2\pm})^{2}},     \nonumber \\ 
M^{B}_{M1\pm} &\equiv&  \sqrt{\frac{c_{2}+c_{3}-|c_{4}|}{2}\mp\frac{1}{2}\sqrt{\bigl( c_{2}-c_{3}+|c_{4}|\bigr)^{2} + 4|c_{1}|^{2}}}  \nonumber \\
M^{B}_{M2\pm} &\equiv&  \sqrt{\frac{c_{2}+c_{3}+|c_{4}|}{2}\mp\frac{1}{2}\sqrt{\bigl( c_{2}-c_{3}-|c_{4}|\bigr)^{2} + 4|c_{1}|^{2}}}  \nonumber \\
c_{1} &\equiv& -m_{D}(g^{\dagger}\phi^{c\dagger}_{X}+2m_{L}), \quad c_{2} \equiv m^{2}_{D} +4m^{2}_{L},  \nonumber \\ 
c_{3} &\equiv& m^{2}_{D} +|g|^{2}(J_{1}+K_{1}+ |\phi^{c}_{+}|^{2} + |\phi^{c}_{X}|^{2}), \quad 
c_{4} \equiv -\frac{|g|^{2}}{4}( J_{2} + (\phi^{c}_{+})^{2})-\frac{fg^{\dagger}}{2}.
\end{eqnarray}
$|c_{1}|^{2}$ includes the phase $\theta_{X}$.
If we put the expressions at the classical solution for $\phi_{+}$, $J_{1}$ and $K_{1}$ to $c_{3}$ and $c_{4}$
with employing the Popov approximation $J_{2}=0$, we get
\begin{eqnarray}
M^{B}_{M1\pm} &=& M^{B}_{M2\pm}   \nonumber \\
&=& \Bigg[ m_{D}^{2} + \frac{|g|^{2}|\phi^{c}_{X}|^{2}}{2} + 2m_{L}^{2} 
+ \frac{m^{4}_{D}}{8m^{2}_{L}} + \frac{m^{3}_{D}}{4m_{L}} -\frac{5}{2}|g|f    \nonumber \\
& & \quad \mp 2 \Bigl\{ 
\Bigl( \frac{|g|^{2}|\phi^{c}_{X}|^{2}}{4}-m_{L}^{2}
+ \frac{m^{4}_{D}}{16m^{2}_{L}} + \frac{m^{3}_{D}}{8m_{L}} -\frac{5}{4}|g|f \Bigr)^{2}
+ m_{D}^{2}\Bigl(\frac{|g|^{2}|\phi^{c}_{X}|^{2}}{4}+m_{L}^{2}+|g||\phi^{c}_{X}|m_{L}\cos\theta_{X}\Bigr) 
\Bigr\}^{1/2}\Bigg]^{1/2}.
\end{eqnarray}
Therefore, we conclude
\begin{eqnarray}
M^{F}_{\pm} &<& M^{B}_{M1\pm}, M^{B}_{M2\pm},
\end{eqnarray}
at $f<0$, $g>0$.
The mass eigenvalues of $M^{B}_{M1+}$ and $M^{B}_{M2+}$ become tachyonic at
\begin{eqnarray}
M^{B}_{M1+}; \, c_{2}(c_{3}-|c_{4}|) < |c_{1}|^{2}, \quad M^{B}_{M2+}; \, c_{2}(c_{3}+|c_{4}|) < |c_{1}|^{2},
\end{eqnarray}
and an appearance of tachyon indicates the instability of vacuum state~[12,45].
In our condition (44) of tachyonic masses, $c_{3}$ and $c_{4}$ include the HFB self-energies.
At the classical solution (7) with $J_{2}=0$, 
the tachyon condition (44) becomes such that
\begin{eqnarray}
2f - g\Bigl(J_{1}+K_{1}\Bigr) = 5f-\frac{1}{g}\Bigl(\frac{m^{4}_{D}}{4m^{2}_{L}}+\frac{m^{3}_{D}}{2m_{L}}\Bigr) > 0.
\end{eqnarray}
Here we have assumed $f,g$ as real.
( Again, we wish to rewrite that $f$, $J_{1}$, $K_{1}$, and $J_{2}$ have mass dimension $[Mass]^{2}$,
while $g$ is dimensionless. )
Hence, the vicinity of the classical solution is stable if $f<0$ and $g>0$.
Due to the HFB self-energies and $\phi^{c}_{+}$,
there are several differences between $M^{F}_{\pm}$ and the mass eigenvalues obtained from $\Omega^{B}_{MM}$:
The mass spectra of bosons and fermions are not symmetric ( namely, not supersymmetric ) in our theory.
We regard the vacuum energy as the order parameter of SUSY-breaking,
we must examine the local/global structure of the one-loop effective potential
to clarify whether the vacuum energy vanishes or not before concluding a breakdown of SUSY.

\vspace{3mm}

In the determinant $\det\{\Omega^{B}_{XX}-\Omega^{B}_{XM}(\Omega^{B}_{MM})^{-1}\Omega^{B}_{MX}\}$,
we wish to concentrate upon the vicinity of the classical solution (7).
At the point (7) with the neglection of $J_{2}$, 
$\Omega^{B}_{++}$ in the expression of (25) becomes diagonal.
This helps us to evaluate the mass eigenvalue of $\phi_{X}$ in an analytic manner.
Then we get $(M^{B}_{X\pm})^{2}$ in terms of bare parameters as follows:
\begin{eqnarray} 
(M^{B}_{X\pm})^{2} &\equiv& \lim_{p^{2}\to 0} (|A|\pm |B|),  \nonumber \\
A &\equiv& -a + \frac{|(|\alpha|^{2}+\beta^{2})b+c|d|^{2}-(\alpha de^{\dagger}+\alpha^{\dagger}d^{\dagger}e)|}{bc-|e|^{2}},  \quad
B \equiv -\frac{|(\beta+\beta^{\dagger})(\alpha^{\dagger}b-de^{\dagger})|}{bc-|e|^{2}},  \nonumber \\
a &\equiv&  - |g|^{2}( J_{1} + |\phi^{c}_{+}|^{2} ),  \quad
b \equiv  \Box - m^{2}_{D} -4m^{2}_{L},  \nonumber \\
c &\equiv&  \Box - m^{2}_{D} -|g|^{2}(J_{1}+K_{1}+ |\phi^{c}_{+}|^{2} + |\phi^{c}_{X}|^{2}),  \quad
d \equiv - g^{\dagger}m_{D}\phi^{c\dagger}_{+}, \quad 
e \equiv - m_{D}(g^{\dagger}\phi^{c\dagger}_{X}+2m_{L}),  \nonumber \\
\alpha &\equiv&  -|g|^{2}( K^{\dagger}_{3} + \phi^{c\dagger}_{+}\phi^{c}_{X} ),   \quad
\beta \equiv -g^{\dagger}m_{D}\phi^{c}_{-} - |g|^{2}( K_{2} + \phi^{c}_{+}\phi^{c}_{X} ).   
\end{eqnarray}
These $a,b,c,d,e,\alpha,\beta$ are matrix elements of $\Omega^{B}$ ( see (25) ).
We also set four-momentum as $p_{\nu}=0$ in the bosonic matrices.
We examine the stability conditions of $M^{B}_{X\pm}$. 
Especially, we have interest on its behavior in the vicinity of the classical solution with the seesaw condition
$|g|^{2}|\phi^{c}_{X}|^{2}\gg m^{2}_{D}\gg m^{2}_{L}$.
A direct evaluation from (46) with (7), (20), (21), and by employing a Popov approximation
$J_{2}=K_{2}=K_{3}=0$ gives
\begin{eqnarray}
(M^{B}_{X+})^{2} &=& (M^{B}_{X-})^{2}   \nonumber \\
&=& -\frac{1}{(m^{2}_{D}+4m^{2}_{L})(3gf+\frac{m^{4}_{D}}{4m^{2}_{L}}+\frac{m^{3}_{D}}{2m_{L}})}  \nonumber \\
& & \qquad \times \Bigl[     
48g^{2}f^{2}m^{2}_{L} + 2g^{2}f^{2}m^{2}_{D} + 8gfm^{3}_{D}m_{L} + 6gfm^{4}_{D}
+ 3gf\frac{m^{5}_{D}}{m_{L}} + \frac{gf}{2}\frac{m^{6}_{D}}{m^{2}_{L}}
\Bigr].
\end{eqnarray}
Hence, if we take into account the seesaw condition $m_{D}\gg m_{L}$, 
then we obtain $g>0$ and $f<0$ is the stability condition of $(M^{B}_{X\pm})^{2}$.
A rough estimation gives
\begin{eqnarray}
(M^{B}_{X\pm})^{2} &\sim& -2gf \sim {\cal O}(gf).
\end{eqnarray}
We conclude that, with taking into account (45) and (48), 
the generalized seesaw mechanism can take place under $f<0$ and $g>0$
( we have assumed that $|gf|(m_{L}/m^{2}_{D})\ll 1$ ).
It is worth noticing that $(M^{B}_{X\pm})^{2}\sim -m^{4}/8m^{2}_{L}$ ( negative ) if we set $K_{1}=0$,
and thus the HFB self-energy $K_{1}$ is important for stability of the potential at the classical solution.
Finally one finds 
\begin{eqnarray}
E^{B}_{X\pm} &=& \sqrt{\bmk^{2}+(M^{B}_{X\pm})^{2}},
\end{eqnarray}
and we obtain the one-loop contribution of the scalar sector as follows:
\begin{eqnarray}
V^{B(1)} &=& \frac{i}{2}{\rm Tr}
(k_{0}-E^{B}_{X+})(k_{0}+E^{B}_{X+})(k_{0}-E^{B}_{X-})(k_{0}+E^{B}_{X-})   \nonumber \\
& & \quad \times (k_{0}-E^{B}_{M1+})(k_{0}+E^{B}_{M1+})(k_{0}-E^{B}_{M1-})(k_{0}+E^{B}_{M1-})    \nonumber \\
& & \quad \times (k_{0}-E^{B}_{M2+})(k_{0}+E^{B}_{M2+})(k_{0}-E^{B}_{M2-})(k_{0}+E^{B}_{M2-}).
\end{eqnarray}

\vspace{3mm}

It is a well-known fact that the naive dimensional regularization, 
suitable to keep a gauge invariance in a non-SUSY gauge theory,
will break SUSY through the regularization. 
To circumvent of this problem is relatively easier in non-gauge models,
while the problem is severe in SUSY gauge theories, 
and the method of "dimensional reduction" regularization seems more suitable~[18].
Since the Lagrangian we consider here is not a gauge model, 
here we employ a simple cutoff scheme for regularizations of integrals.
After performing the four-dimensional momentum integration, 
the one-loop contribution to the effective potential will be obtained as follows:
\begin{eqnarray}
V^{(1)} &=& \frac{1}{16\pi^{2}}\Bigg[ \Lambda^{2}\Bigl\{M^{B2}_{X+}+M^{B2}_{X-}+M^{B2}_{M1+}+M^{B2}_{M1-}+M^{B2}_{M2+}+M^{B2}_{M2-}-2(M^{F2}_{X}+M^{F2}_{+}+M^{F2}_{-})\Bigr\}   \nonumber \\
& & +\Lambda^{4}\ln\frac{(1+M^{B2}_{X+}/\Lambda^{2})(1+M^{B2}_{X-}/\Lambda^{2})(1+M^{B2}_{M1+}/\Lambda^{2})(1+M^{B2}_{M1-}/\Lambda^{2})(1+M^{B2}_{M2+}/\Lambda^{2})(1+M^{B2}_{M2-}/\Lambda^{2})}
{(1+M^{F2}_{X}/\Lambda^{2})^{2}(1+M^{F2}_{+}/\Lambda^{2})^{2}(1+M^{F2}_{-}/\Lambda^{2})^{2}}    \nonumber \\
& & -M^{B4}_{X+}\ln\Bigr(1+\frac{\Lambda^{2}}{M^{B2}_{X+}}\Bigl)-M^{B4}_{X-}\ln\Bigr(1+\frac{\Lambda^{2}}{M^{B2}_{X-}}\Bigl) +2M^{F4}_{X}\ln\Bigl(1+\frac{\Lambda^{2}}{M^{F2}_{X}}\Bigr) 
+2M^{F4}_{+}\ln\Bigl(1+\frac{\Lambda^{2}}{M^{F2}_{+}}\Bigr) + 2M^{F4}_{-}\ln\Bigl(1+\frac{\Lambda^{2}}{M^{F2}_{-}}\Bigr)   \nonumber \\
& & -M^{B4}_{M1+}\ln\Bigr(1+\frac{\Lambda^{2}}{M^{B2}_{M1+}}\Bigl) -M^{B4}_{M1-}\ln\Bigl(1+\frac{\Lambda^{2}}{M^{B2}_{M1-}}\Bigr) 
-M^{B4}_{M2+}\ln\Bigr(1+\frac{\Lambda^{2}}{M^{B2}_{M2+}}\Bigl) -M^{B4}_{M2-}\ln\Bigl(1+\frac{\Lambda^{2}}{M^{B2}_{M2-}}\Bigr) \Bigg],
\end{eqnarray}
where, $\Lambda$ denotes the four-momentum cutoff. 
By the standard method for handling the effective potential, 
namely, remove contributions they will vanish at $\Lambda\to\infty$ 
( there is no divergent constant due to ${\cal N}=1$ SUSY ), one obtains
\begin{eqnarray}
V^{(1)} &=& \frac{1}{16\pi^{2}}\Bigg[ 
\bigl(M^{B}_{X+}\bigr)^{4}\ln\Bigl(\frac{M^{B}_{X+}}{\Lambda}\Bigr)^{2} +\bigl(M^{B}_{X-}\bigr)^{4}\ln\Bigl(\frac{M^{B}_{X-}}{\Lambda}\Bigr)^{2}
- 2\bigl(M^{F}_{X}\bigr)^{4}\ln\Bigl(\frac{M^{F}_{X}}{\Lambda}\Bigr)^{2} 
-2\bigl(M^{F}_{+}\bigr)^{4}\ln\Bigl(\frac{M^{F}_{+}}{\Lambda}\Bigr)^{2} - 2\bigl(M^{F}_{-}\bigr)^{4}\ln\Bigl(\frac{M^{F}_{-}}{\Lambda}\Bigr)^{2}  \nonumber \\
& & \quad +\bigl(M^{B}_{M1+}\bigr)^{4}\ln\Bigl(\frac{M^{B}_{M1+}}{\Lambda}\Bigr)^{2} +\bigl(M^{B}_{M1-}\bigr)^{4}\ln\Bigl(\frac{M^{B}_{M1-}}{\Lambda}\Bigr)^{2} 
+\bigl(M^{B}_{M2+}\bigr)^{4}\ln\Bigl(\frac{M^{B}_{M2+}}{\Lambda}\Bigr)^{2} +\bigl(M^{B}_{M2-}\bigr)^{4}\ln\Bigl(\frac{M^{B}_{M2-}}{\Lambda}\Bigr)^{2} \Bigg],   \nonumber \\
& & ( \, M^{B}_{X\pm},M^{F}_{X},M^{B}_{M1\pm},M^{B}_{M2\pm},M^{F}_{\pm}\ll \Lambda \, ).
\end{eqnarray} 
We have arrived at a generalization of the so-called SUSY Coleman-Weinberg potential discussed in Refs.~[11,12]
( see also, Ref.~[46] ).
In our $V^{(1)}$, the one-loop contribution of $X$-field is also included.
For obtaining a one-loop potential which will not diverge 
into the negative-energy direction at the limit $|\phi^{c}_{X}|\to\infty$~[47], 
we should impose both
\begin{eqnarray}
(M^{F}_{\pm}) < (M^{B}_{M1\pm})^{2}, (M^{B}_{M2\pm})^{2}
\end{eqnarray}
and
\begin{eqnarray}
(M^{F}_{X})^{2} < (M^{B}_{X\pm})^{2}.
\end{eqnarray}
We have known from (43) that (53) is satisfied, while if
\begin{eqnarray}
& & \theta_{X} \sim \pi, \quad 1 > g > 0, \quad 0 > f(m_{L}/m^{2}_{D}) > -1, 
\end{eqnarray}
then (54) is satisfied.
We should set model parameters with respect to these relations. 
The mass eigenvalues will degenerate under using (7) for $\phi^{c}_{\pm}$ with the Popov approximation.
Therefore, we will denote them as follows:
\begin{eqnarray}
M^{B}_{X} \equiv M^{B}_{X+} = M^{B}_{X-}, \quad
M^{B}_{+} \equiv M^{B}_{M1+} = M^{B}_{M2+}, \quad
M^{B}_{-} \equiv M^{B}_{M1-} = M^{B}_{M2-}.
\end{eqnarray}
Since $m_{D},f\gg m_{L}$, especially we have interest on a situation 
\begin{eqnarray}
M^{F}_{X} \ll M^{B}_{X} < M^{F}_{+} < M^{B}_{+} \ll M^{F}_{-} < M^{B}_{-}.
\end{eqnarray} 
We have arrived at the order of mass eigenvalues and it is the crucial result for our discussion hereafter.

\vspace{3mm}

Since our effective potential and mass eigenvalues have similarities with 
those of the Minimal Supersymmetric Standard Model ( MSSM )~[19,20],
let us utilize some methods/results from it.
In theory of MSSM, SUSY is expricitly broken by a vacuum energy and several soft mass parameters,
while it is not broken in the Lagrangian level as the starting point of our model.
Thus, discussion on renormalization would become simpler than that of MSSM.
A renormalization-group invariant calculation for renormalization of our $V^{tree}+V^{(1)}$ is subtle,
because it includes many different mass parameters/scales~[19,20,48-53].
Since our interest is to examine a possibility of a realization of the generalized seesaw mechanism
in the vicinity of the classical solution (7) with taking into account the one-loop contribution (52),
we concentrate on a VEV of $\phi_{X}$ under the situation (55). 
Unfortunately, it is difficult to find a global minimum of the potential $V^{tree}+V^{(1)}$
because it has many parameters, $\phi^{c}_{\pm}$, $\phi^{c}_{X}$, $J_{1}$ and $K_{1}$ which should be determined variationally. 
For example, if we put the expressions of classical solution (7) for condensate $\phi^{c}_{+}$
to reduce variational parameters and try to find a minimum with respect to variation of $\phi^{c}_{X}$,
the potential might give a non-vanishing vacuum energy ( hence SUSY is broken ) 
because this procedure corresponds to a restriction of trial functions in the variation:
To achieve a true vacuum might be difficult. 
In the usual prescription of renomalization, 
a running coupling will be used to remove a renormalization point 
from a theory to get a physical ( renormalization-group invariant ) potential,
though this procedure is difficult in our case.
To make our problem tractable for our purpose,
we will use the following definition of $V^{(1)}$~[50,53] 
by taking into account the Appelquist-Carazzone decoupling theorem~[54]:
\begin{eqnarray}
V^{(1)} &=& \frac{1}{8\pi^{2}}\sum_{l}\Bigl[ \theta(\mu^{2}-(M^{B}_{l})^{2})(M^{B}_{l})^{4}\ln\frac{(M^{B}_{l})^{2}}{\mu^{2}} 
- \theta(\mu^{2}-(M^{F}_{l})^{2})(M^{F}_{l})^{4}\ln\frac{(M^{F}_{l})^{2}}{\mu^{2}} \Bigr],  \nonumber \\
& & ( l = X, +, - ).
\end{eqnarray}
( A quite clear example of the decoupling theorem can be found in Ref.~[20]. )
Here, $\theta(x)$ is the Heaviside step function,
it has been introduced to define mass thresholds inside the potential. 
We have changed the regularization method to $\overline{MS}$ ( modified minimal subtraction scheme ).
$\mu\equiv e^{3/2}\bar{\mu}$, where $\bar{\mu}$ denotes the $\overline{MS}$ renormalization scale.
Of course, $M^{B}_{l}$ and $M^{F}_{l}$ are functions of $\phi^{c}_{X}$.
The logarithmic functions appeared in the above equation must satifiy
$|\ln(M^{2}/\mu^{2})| < 1$ ( $\mu$: a renomalization point  ) for the justification for our loop expansion.
We should find the situation where $V^{(1)}(\mu)=0$ and $\frac{d}{d\mu}(V^{tree}+V^{(1)})=0$
are simultaneously satisfied~[50].
It is a hard task to arrive from the complete theory to an effective theory of lowest region in (57) 
with running parameters with $\mu$. 
In such a top-down approach, as we know from (57), we have totally six decoupling scales until we arrive at
the region $\mu^{2}<(M^{F}_{X})^{2}$ where all particles are decoupled, 
and then $V^{tree}$ alone gives the renormalization-group invariant potential~[50,53].
Hence, first we wish to consider the case $(M^{F}_{X})^{2} < \mu^{2} < {\rm others}$.
In this case, the potential of the one-loop contribution can be written down as follows:
\begin{eqnarray}
V^{(1)} &=& -\frac{1}{8\pi^{2}}(M^{F}_{X})^{4}\ln\frac{(M^{F}_{X})^{2}}{\mu^{2}}. 
\end{eqnarray}
Here, we simply have assumed that the effect of decoupled particles 
is already included by a renormalization of parameter.
This $V^{(1)}$ gives a positive contribution to our one-loop potential.
After put the classical solution (7) to $\phi^{c}_{\pm}$ of this $V^{(1)}$, 
choose $\theta_{X}=\pi$, 
and take the derivative of $V^{tree}+V^{(1)}$ with respect to $|\phi_{X}|$,
we get $\langle\phi^{c}_{X}\rangle = m^{2}_{D}/(2gm_{L})$ from the stationary condition: 
We find that the effective field theory of this renormalization point/scale gives 
the same expression for VEV of $\phi_{X}$ with its classical solution,
shows the generalized seesaw mechanism.
Needless to say, we will also obtain $\langle\phi^{c}_{X}\rangle = m^{2}_{D}/(2gm_{L})$ at the
complete decoupled region $\mu^{2}<(M^{F}_{X})^{2}$ because $V^{(1)}=0$.
Since the improved $V^{tree}$ is the "exact" potential ( with satisfying the matching condition )~[50,53],
an observation will find that the vacuum is supersymmetric in the energy scale $\mu^{2}<(M^{F}_{X})^{2}$.
We conclude that, under a reasonable choice of model parameters with respect to the conditions for stability
of the vicinity of the classical solution (7) and a justfication on convergence of the loop expansion,
certainly (7) is robust against a quantum correction, will not obtain a radical modification,
and thus the generalized seesaw mechanism takes place.

\section{Superspace Formalism}

In this section we will calculate the one-loop effective potential in superfield formalism~[2,17,55],
though we have to use components of superfields at several points of our discussions, especially for our consideration on BEC.
For example, it seems difficult to consider the HFB approximation of quantum fluctuations of scalars in the superfield formalism,
and thus the self-energies as $J_{1}, K_{1}, \cdots$ do not appear in our superspace formalism.
It is a problem inherent in the superspace formalism, 
and as a result, the one-loop contribution of the superspace formalism is different from that of the component field formalism.
Moreover, it seems difficult to examine the generalized seesaw mechanism by our superfield formalism of
one-loop potential because mass eigenvalues of fermion and boson sectors will not be derived under a direct manner.
Therefore the component field formalism is better to describe the dynamics and physical property of the scalar sector with having the BEC.
The purpose of this section is to make a comparison between the two formalisms. 
The examination on the generalized seesaw mechanism is beyond scope of this section.
We employ the background field method, the standard method of superspace formalism~[17], 
to take into account the BEC in our model:
\begin{eqnarray}
& & X = X^{c} + \tilde{X}, \quad 
X^{c} \equiv \phi^{c}_{X} + \theta\theta F^{c}_{X}, \quad
\tilde{X} \equiv \tilde{\phi}_{X} + \theta \psi_{X} + \theta\theta\tilde{F}_{X},   \nonumber \\ 
& & \Phi_{\pm} = \Phi^{c}_{\pm} + \tilde{\Phi}_{\pm}. \quad 
\Phi^{c}_{\pm} \equiv \phi^{c}_{\pm} + \theta\theta F^{c}_{\pm}, \quad
\tilde{\Phi}_{\pm} \equiv \tilde{\phi}_{\pm} + \theta \psi_{\pm} + \theta\theta\tilde{F}_{X}.
\end{eqnarray}
Similar to the case of component field formalism given in the previous section,
again we assume the condensates $\phi^{c}_{X}$ and $\phi^{c}_{\pm}$ are independent on spacetime coordinates.
The Lagrangian will be converted into the following form: 
\begin{eqnarray}
{\cal L} &=& {\cal L}^{c} + \widetilde{\cal L},  \nonumber \\
{\cal L}^{c} &\equiv& \Bigl( X^{c\dagger}X^{c} + \Phi^{c\dagger}_{+}\Phi^{c}_{+} + \Phi^{c\dagger}_{-}\Phi^{c}_{-} \Bigr)\Big|_{\theta\theta\bar{\theta}\bar{\theta}} 
+ \Bigg[ \Bigl( fX^{c} + \frac{g}{2}X^{c}\Phi^{c}_{+}\Phi^{c}_{+} + m_{D}\Phi^{c}_{+}\Phi^{c}_{-} + m_{L}\Phi^{c}_{-}\Phi^{c}_{-} \Bigr)\Big|_{\theta\theta} + ({\rm h.c.}) \Bigg],    \nonumber \\
\widetilde{\cal L} &\equiv& 
\Bigg[ \tilde{X}^{\dagger}\tilde{X} +\frac{1}{2}\Xi^{\dagger}{\cal M}\Xi 
-\frac{D^{2}}{4\Box} g\Phi^{c}_{+}\tilde{X}\tilde{\Phi}_{+} 
-\frac{\overline{D}^{2}}{4\Box} g^{\dagger}\Phi^{c\dagger}_{+}\tilde{X}^{\dagger}\tilde{\Phi}^{\dagger}_{+}
\Bigg]\Bigg|_{\theta\theta\bar{\theta}\bar{\theta}}. 
\end{eqnarray}
Here, we have used the equivalent relations $\delta^{2}(\bar{\theta})=-D^{2}/4\Box$ and $\delta^{2}(\theta)=-\overline{D}^{2}/4\Box$ under the integration of $d^{4}x$ inside
the action functional of the theory, and have dropped terms linear in the fluctuating ${\it superfields}$
$\tilde{X}$, $\tilde{\Phi}_{\pm}$.
We have introduced several matrix notations defined as follows:
\begin{eqnarray}
\Xi &\equiv& (\tilde{\Phi}_{+},\tilde{\Phi}_{-},\tilde{\Phi}^{\dagger}_{+},\tilde{\Phi}^{\dagger}_{-})^{T},  \nonumber \\
{\cal M} &\equiv& \left( 
\begin{array}{cc}
-\frac{D^{2}\overline{D}^{2}}{16}\otimes\sigma_{0}  & -\frac{D^{2}}{4}{\cal C}^{\dagger}  \\
-\frac{\overline{D}^{2}}{4}{\cal C}  & -\frac{\overline{D}^{2}D^{2}}{16}\otimes\sigma_{0}
\end{array}
\right)\delta^{8}(z-z'),  \quad
{\cal C} \equiv m_{D}\otimes\sigma_{1}+m_{L}\otimes\frac{1-\sigma_{3}}{2}+\frac{g}{2}X^{c}\otimes\frac{1+\sigma_{3}}{2}. 
\end{eqnarray}
The sigma matrices $\sigma^{\nu}$ ( the definition: $\sigma^{0}=-1_{2\times 2}$, while $\sigma^{1},\sigma^{2},\sigma^{3}$ are the ordinary Pauli matrices ) 
act on the two-dimensional chirality space $(+,-)$.
The chiral and antichiral delta functions are defined as
\begin{eqnarray}
\frac{\delta \Phi_{\pm}(z')}{\delta \Phi_{\pm}(z)} = -\frac{\overline{D}^{2}}{4}\delta^{8}(z-z'), \quad 
\frac{\delta \Phi^{\dagger}_{\pm}(z')}{\delta \Phi^{\dagger}_{\pm}(z)} = -\frac{D^{2}}{4}\delta^{8}(z-z'), \quad z \equiv (x,\theta,\bar{\theta}).
\end{eqnarray}
The generating functional will be written down in the following form:
\begin{eqnarray}
{\cal Z} &=& \int {\cal D}\tilde{X}{\cal D}\tilde{X}^{\dagger}{\cal D}\tilde{\Phi}_{+}{\cal D}\tilde{\Phi}^{\dagger}_{+}{\cal D}\tilde{\Phi}_{-}{\cal D}\tilde{\Phi}^{\dagger}_{-}\exp\Bigl[i\int d^{4}x{\cal L} + ({\rm sources}) \Bigr]    \nonumber \\
&=& \int {\cal D}\tilde{X}{\cal D}\tilde{X}^{\dagger}
\exp\Bigg(i \int d^{4}x \Bigl[ {\cal L}^{c} + \Bigl( \tilde{X}^{\dagger}\tilde{X} \Bigr)_{\theta^{2}\bar{\theta}^{2}} \Bigr] + \frac{i}{2}{\rm Tr}\ln{\cal M} + {\cal G} \Bigg),   \nonumber \\
{\cal G} &\equiv& -\frac{i}{2}\int d^{8}z\int d^{8}z' \frac{1}{2}{\cal J}(z){\cal M}^{-1}\delta^{8}(z-z'){\cal J}^{\dagger}(z'). 
\end{eqnarray}
Here, $d^{8}z \equiv d^{4}x d^{2}\theta d^{2}\bar{\theta}$.
To obtain the final expression of ${\cal Z}$ in (64), 
we have neglected contributions of (anti)chiral sources. 
The definition of ${\cal J}$ is
\begin{eqnarray}
{\cal J} &\equiv& \left(
\begin{array}{cc}
g\Phi^{c}_{+}\tilde{X}\otimes\frac{1+\sigma_{3}}{2} & 0 \\
0 & g^{\dagger}\Phi^{c\dagger}_{+}\tilde{X}^{\dagger}\otimes\frac{1+\sigma_{3}}{2}
\end{array}
\right).
\end{eqnarray}
By putting the components of $X^{c}$ and $\Phi^{c}_{\pm}$, we can confirm the fact that ${\cal L}^{c}$ becomes
\begin{eqnarray}
{\cal L}^{c} &=& -V^{tree}[\phi^{c}_{\pm},\phi^{c}_{X}].
\end{eqnarray}
Hence the tree level potential is the same in both of the formalisms. 
Next, we divide ${\cal M}$ as follows:
\begin{eqnarray}
& & {\cal M} = {\cal M}_{0} - {\cal M}', \quad 
{\cal M}_{0} \equiv \left(
\begin{array}{cc}
-\frac{D^{2}\overline{D}^{2}}{16}\otimes\sigma_{0}  & 0  \\
0 & -\frac{\overline{D}^{2}D^{2}}{16}\otimes\sigma_{0}
\end{array}
\right),  \nonumber \\
& & {\cal M}^{-1}_{0} = \frac{1}{\Box^{2}}\left(
\begin{array}{cc}
-\frac{D^{2}\overline{D}^{2}}{16}\otimes\sigma_{0}  & 0  \\
0 & -\frac{\overline{D}^{2}D^{2}}{16}\otimes\sigma_{0}
\end{array}
\right),  \quad
{\cal M}' \equiv \left(
\begin{array}{cc}
0  & +\frac{D^{2}}{4}{\cal C}^{\dagger}  \\
+\frac{\overline{D}^{2}}{4}{\cal C}  & 0
\end{array}
\right). 
\end{eqnarray}
The one-loop effective action $\frac{i}{2}{\rm Tr}\ln{\cal M}$ is evaluated to be
\begin{eqnarray}
\Gamma^{(1)}_{(super)} &\equiv& \frac{i}{2}{\rm Tr}\ln{\cal M}  
= \frac{i}{2}\ln{\rm Det}{\cal M}_{0} + \frac{i}{2}{\rm Tr}\ln(1-{\cal M}^{-1}_{0}{\cal M}') 
= \frac{i}{2}{\rm Tr}\ln\Bigl( 1 -\frac{1}{\Box}{\cal M}'\Bigr)\Box{\cal M}^{-1}_{0}   \nonumber \\
&=& \lim_{z'\to z}\frac{i}{2}{\rm tr}\int d^{8}z \ln\Bigl[ 1 -\frac{1}{\Box}{\cal C}^{\dagger}{\cal C} \Bigr]\frac{D^{2}\overline{D}^{2}}{16\Box}\delta^{8}(z-z').
\end{eqnarray}
We have dropped $\frac{i}{2}\ln{\rm Det}{\cal M}^{-1}_{0}$ because it does not contribute to $\Gamma^{(1)}_{(super)}$.
The relations ${\cal M}^{-1}_{0}{\cal M}^{-1}_{0}=\Box^{-1}{\cal M}^{-1}_{0}$ and the commutator $[{\cal M}^{-1}_{0},{\cal M}']=0$ have been used.
From the following identity in superspace,
\begin{eqnarray}
\frac{D^{2}\overline{D}^{2}}{16}\delta^{2}(\theta-\theta')\delta^{2}(\bar{\theta}-\bar{\theta}')\Big|_{\theta=\theta',\bar{\theta}=\bar{\theta}'} = 1,
\end{eqnarray}
the effective potential is found to be
\begin{eqnarray}
V^{(1)}_{(super)} &\equiv& -\frac{\Gamma^{(1)}_{(super)}}{\int d^{4}x} 
= \frac{i}{2}{\rm tr}\int d^{2}\theta d^{2}\bar{\theta}\int\frac{d^{4}p}{(2\pi)^{4}}\frac{1}{p^{2}}\ln\Bigl(p^{2}+{\cal C}^{\dagger}{\cal C}\Bigr)   \nonumber \\
&=& \frac{1}{2}{\rm tr}\Bigg[ \Lambda^{2}\ln\Bigl(1+\frac{{\cal C}^{\dagger}{\cal C}}{\Lambda^{2}}\Bigr)
+{\cal C}^{\dagger}{\cal C}\ln\Bigl(1+\frac{\Lambda^{2}}{{\cal C}^{\dagger}{\cal C}}\Bigr)\Bigg]_{\theta^{2}\bar{\theta}^{2}}     \nonumber \\
&=& \frac{|g|^{2}}{4}|F^{c}_{X}|^{2}\ln\Bigl(1 + \frac{\Lambda^{2}}{|M_{\cal C}|^{2}} \Bigr) - \Bigl[\frac{|g|^{2}}{4}\Bigr]^{2}|F^{c}_{X}|^{2}|\phi^{c}_{X}|^{2}\frac{\Lambda^{2}}{(\Lambda^{2}+|M_{\cal C}|^{2})|M_{\cal C}|^{2}}   \nonumber \\
&\approx& |F^{c}_{X}|^{2}\frac{|g|^{2}}{4}\ln \frac{\Lambda^{2}}{|M_{\cal C}|^{2}},  \quad ( \, \Lambda \to \infty \,  ),  \nonumber \\
|M_{\cal C}|^{2} &=& m^{2}_{D} + m^{2}_{L} + \frac{|g|^{2}}{4}|\phi^{c}_{X}|^{2}.
\end{eqnarray}
Of course, the mass dimension of $V^{(1)}_{(super)}$ is $[{\rm mass}]^{4}$. 
Both $V^{tree}$ and $V^{(1)}_{(super)}$ vanish simultaneously at the classical vacuum (7) and SUSY is not broken.
To make our calculation on ${\cal G}$ in (64) tractable, 
we approximate ${\cal M}^{-1}$ by replacing ${\cal C}\to m_{D}$.
Then we get
\begin{eqnarray}
{\cal G} &=& -\frac{i}{4} \Bigg[ \int d^{8}z |g|^{2}|\Phi^{c}_{+}|^{2}\Bigl( \tilde{X}\frac{1}{\Box-m^{2}_{D}}\tilde{X}^{\dagger} + \tilde{X}^{\dagger}\frac{1}{\Box-m^{2}_{D}}\tilde{X}\Bigr)    \nonumber \\
& & \quad + \int d^{6}z (g)^{2}(\Phi^{c}_{+})^{2}\Bigl( \tilde{X}\frac{m_{D}}{\Box-m^{2}_{D}}\tilde{X} \Bigr) + \int d^{6}\bar{z} (g^{\dagger})^{2}(\Phi^{c\dagger}_{+})^{2}\Bigl( \tilde{X}^{\dagger}\frac{m_{D}}{\Box-m^{2}_{D}}\tilde{X}^{\dagger} \Bigr) \Bigg].
\end{eqnarray}
Obviously, ${\cal G}$ includes a K\"{a}hler potential and (anti)chiral superpotentials of the fluctuating $\tilde{X}$-field. 
From a consideration by the Wick theorem, one finds $(\Box-m^{2}_{D})^{-1}$ in the K\"{a}hler potential corresponds to 
the propagator $\langle T\tilde{\phi}_{+}\tilde{\phi}^{\dagger}_{+}\rangle$, 
while $m_{D}/(\Box-m^{2}_{D})$ in the chiral and antichiral superpotential parts of ${\cal G}$ came 
from $\langle T\tilde{F}_{+}\tilde{\phi}_{+}\rangle$ and $\langle T\tilde{F}^{\dagger}_{+}\tilde{\phi}^{\dagger}_{+}\rangle$, 
respectively.
( An examination of mass dimensions of these propagators is also helpful. ) 
Because $(\Phi^{c}_{+})^{2}\tilde{X}$ or $(1/(\Box -m^{2}_{D}))\tilde{X}$ are chiral superfields,
$-\frac{D^{2}}{4\Box}$ can be inserted between them inside the integration $\int d^{8}z$.
Therefore we get
\begin{eqnarray}
\int d^{8}z \tilde{X}^{\dagger}\tilde{X} -i{\cal G} &=& \frac{1}{2}\int d^{8}z (X^{\dagger},X){\cal M}_{X}\left(
\begin{array}{c}
X \\
X^{\dagger}
\end{array}
\right),  \nonumber \\
{\cal M}_{X} &\equiv& \left(
\begin{array}{cc}
1-\frac{g^{2}}{2}|\Phi^{c}_{+}|^{2}\frac{1}{\Box -m^{2}_{D}} & -\frac{g^{2}}{2}(\Phi^{c\dagger}_{+})^{2}(-\frac{\overline{D}}{4\Box})\frac{m_{D}}{\Box -m^{2}_{D}}  \\
-\frac{g^{2}}{2}(\Phi^{c}_{+})^{2}(-\frac{D}{4\Box})\frac{m_{D}}{\Box -m^{2}_{D}} & 1-\frac{g^{2}}{2}|\Phi^{c}_{+}|^{2}\frac{1}{\Box -m^{2}_{D}}
\end{array}
\right).
\end{eqnarray}
Integration of ${\cal D}\tilde{X}{\cal D}\tilde{X}^{\dagger}$ will give ${\rm Det}^{-1}{\cal M}_{X}$,
and this determinant gives a polynomial of $F^{c}_{+}$ and $F^{c\dagger}_{+}$.
Because $F^{c}_{+}=F^{c\dagger}_{+}=0$ at (7), 
the one-loop contribution of ${\rm Det}^{-1}{\cal M}_{X}$ is also vanish 
and we conclude that SUSY is not broken at the classical solution (7).

\section{Conclusion}

In summary, we have examined the mass spectra of scalars and spinors of the modified O'Raifeartaigh model 
by our evaluation of the one-loop effective potential in the component field formalism,
especially in the vicinity of the classical solution (7) of the model,
from the context of the generalized seesaw mechanism.
The BEC in the scalar sector has been considered, while the spinor sector has a mathematical similarity
with relativistic theory of superconductivity~[56,57].
Therefore, some parts of our formulation has some similarities with that of 
theory of supersymmetric (color-)superconductivity~[58,59],
though the intrinsic dynamics of them are quite different.
We have emphasized that it becomes possible for us to examine the mass spectra for the generalized seesaw mechanism
of neutrino by introducing the left-handed Majorana mass term ( namely, a modification~[11,12] )
to the ordinary O'Raifeartaigh model.
Our calculation at the one-loop level of the effective potential of the component field formalism 
indicates that SUSY is not broken in the theory due to the absence of an NG fermion,
and have confirmed that SUSY is not broken at the classical vacuum 
of the one-loop potential of the superfield formalism.

\vspace{3mm}

In this paper, we have discussed several VEVs of scalars. 
It is interesting for us to consider some possible relations between inflaton of cosmology,
scalar fields of (1), a ( generalized ) seesaw mechanism of neutrino,
and a spontaneous SUSY breaking.
The scalar field $\phi_{X}$ seems to have a special role 
in a determination of local/global minima of an effective potential of (1), 
while its VEV determines a right-handed Majorana mass parameter in our theory. 
Thus it is interesting for us to investigate a possible scenario of a relation 
between $\phi_{X}$ and an inflaton for our further investigation.
It might be possible to investigate the lightest fermion $\psi_{X}$-field could become 
a candidate of dark matter.
In the strong CP problem of QCD, a theta angle gives us an important issue on axion.
An investigation of a relation between $\theta_{X}$ and QCD theta angle 
( and also the Peccei-Quinn mechanism~[60] ) is far beyond scope of this paper,
though it is also an interesting problem.

\end{document}